\documentclass[preprint,5p,times,twocolumn]{elsarticle}

\usepackage{graphicx}
\usepackage{amsmath,amsthm,amssymb,amsfonts}
\usepackage{units}
\usepackage{textcomp}
\usepackage{silence}
\usepackage{doi}

\begin{document}

\begin{frontmatter}
\title{Constraining interactions mediated by axion-like particles with ultracold neutrons}


\author[ETH,PSI,JUH]{S.~Afach}
\author[LPC]{G.~Ban}
\author[PSI]{G.~Bison}
\author[JUC]{K.~Bodek} 
\author[PTB]{M.~Burghoff} 
\author[PSI]{M.~Daum}       
\author[ETH,PSI]{M.~Fertl\fnref{MF}}		
\author[ETH,PSI]{B.~Franke\fnref{BF}}   \ead{beatrice.franke@mpq.mpg.de}
\author[UNIFR]{Z.~D.~Gruji\'c}
\author[PSI,LPC]{V.~H\'elaine}   	
\author[UNIFR]{M.~Kasprzak}
\author[LPSC]{Y.~Kerma\"{i}dic}
\author[ETH,PSI]{K.~Kirch}
\author[UNIFR]{P. Knowles\fnref{PK}}
\author[UNIFR,PGUM]{H.-C.~Koch}
\author[ETH,PSI]{S.~Komposch}
\author[HNINP]{A.~Kozela}
\author[ETH]{J.~Krempel} 
\author[PSI]{B.~Lauss}     \ead{bernhard.lauss@psi.ch}    	
\author[LPC]{T.~Lefort}       	
\author[LPC]{Y.~Lemi\`ere}  
\author[PSI]{A.~Mtchedlishvili}      
\author[LPC]{O.~Naviliat-Cuncic\fnref{ONC}}
\author[ETH]{F.~M.~Piegsa}	
\author[LPSC]{G.~Pignol}       
\author[KULEUVEN]{P.~N.~Prashanth}
\author[LPC]{G.~Qu\'em\'ener} 	
\author[LPSC]{D.~Rebreyend}
\author[ETH,PSI]{D.~Ries}    	
\author[CSNSM]{S.~Roccia}  \ead{stephanie.roccia@csnsm.in2p3.fr}
\author[PSI]{P.~Schmidt-Wellenburg}
\author[PTB]{A.~Schnabel}  
\author[KULEUVEN]{N.~Severijns}    
\author[PTB]{J.~Voigt}		
\author[UNIFR]{A. Weis} 
\author[ETH,JUC]{G.~Wyszynski}        	
\author[JUC]{J.~Zejma}        
\author[ETH]{J.~Zenner}		
\author[PSI]{G.~Zsigmond}     	

\address[ETH]{ETH Z\"urich, Institute for Particle Physics, CH-8093 Z\"urich, Switzerland}
\address[PSI]{Paul Scherrer Institute (PSI), CH--5232 Villigen-PSI, Switzerland}
\address[JUH]{Hans Berger Department of Neurology, Jena University Hospital, D-07747 Jena, Germany}
\address[LPC]{LPC Caen, ENSICAEN, Universit\'e de Caen, CNRS/IN2P3, Caen, France}
\address[JUC]{Marian Smoluchowski Institute of Physics, Jagiellonian University, 30--059 Cracow, Poland}
\address[PTB]{Physikalisch Technische Bundesanstalt, Berlin, Germany}
\address[UNIFR]{Physics Department, University of Fribourg, CH-1700 Fribourg, Switzerland}
\address[LPSC]{LPSC, Universit\'e  Grenoble Alpes, CNRS/IN2P3, Grenoble, France}
\address[PGUM]{Institut f\"ur Physik, Johannes--Gutenberg--Universit\"at, D--55128 Mainz, Germany}
\address[HNINP]{Henryk Niedwodnicza\'nski Institute for Nuclear Physics, 31--342 Cracow, Poland}
\address[KULEUVEN]{Instituut voor Kern-- en Stralingsfysica, University of Leuven, B--3001 Leuven, Belgium}
\address[CSNSM]{CSNSM, Universit\'e Paris Sud, CNRS/IN2P3, Orsay Campus, France}

\fntext[MF]{Now at University of Washington, Seattle WA, USA.}
\fntext[BF]{Now at Max-Planck-Institute of Quantum Optics, Garching, Germany.}
\fntext[PK]{Now at LogrusData,Vienna, Austria}
\fntext[ONC]{Now at Michigan State University, East-Lansing, USA.}


\begin{abstract}
We report
a new limit on a possible short range spin-dependent interaction 
from the precise measurement of the ratio of 
Larmor precession frequencies of stored ultracold neutrons and $^{199}$Hg atoms 
confined in the same volume.
The measurement was performed in a $\sim$1\,\textmu T vertical magnetic holding field
with the apparatus searching for a permanent electric dipole moment of the neutron at the Paul Scherrer Institute. 
A possible coupling between freely precessing polarized neutron spins 
and unpolarized nucleons of the wall material can be investigated by searching for a tiny change of the precession frequencies of neutron and mercury spins.
Such a frequency change 
can be interpreted as a consequence of a short range spin-dependent interaction that could possibly be mediated by axions or axion-like particles.
The interaction strength is proportional to the CP violating product of scalar and pseudoscalar coupling constants $g_Sg_P$.
Our result confirms limits from complementary experiments with spin-polarized nuclei in a model-independent way.
Limits from other neutron experiments are improved by up to two orders of magnitude in the interaction range of
$10^{-6}<\lambda<10^{-4}$\,m.
\end{abstract}
\end{frontmatter}


\section{Introduction}
\label{sec:intro}

We present an interpretation of our recent measurement of the ratio $\gamma_\mathrm{n}/\gamma_\mathrm{Hg}$ of the neutron and $^{199}$Hg magnetic moments \cite{pignol2014} in terms of the strength of a possible short range spin-dependent neutron-nucleon interaction.
This ratio was inferred from a comparison of the simultaneously recorded Larmor precession frequencies of the two species contained in the same storage volume. 
The measurement was performed using the apparatus dedicated to the search for the neutron electric dipole moment (nEDM) 
\cite{nEDMatPSI2011} by the nEDM collaboration 
at the UCN source \cite{psiucnsource} of the Paul Scherrer Institute, Switzerland.

In the central storage vessel of the apparatus,
the spins of the neutrons and mercury atoms are made to precess simultaneously in the same volume.
The ratio
\begin{equation}
R=\frac{f_\mathrm{n}}{f_\mathrm{Hg}}
\label{eq:ratioR}
\end{equation}
constitutes a sensitive 
tool for the control of systematic effects during the measurement of the nEDM.
By correcting $R$ properly for known differences of the Larmor precession of the two species UCN and $^{199}$Hg, respectively, the ratio of magnetic moments $\gamma_\mathrm{n}/\gamma_\mathrm{Hg}$ can be extracted.
A data set of $R$ taken in 2012 was independently analysed 
in \cite{pignol2014} 
and 
in \cite{Franke2013}, 
where we additionally examined its sensitivity to hypothetical short range spin-dependent interactions.
Possible force mediators are axions, or axion-like particles and the interaction strength is proportional to the product of scalar and pseudoscalar coupling constants $g_Sg_P$.
It has been proposed in \cite{Zimmer2010physB,Serebrov2010Axion} to use an nEDM apparatus for the investigation of such a force.

A motivation to search for an interaction involving $g_Sg_P$ is given in Sec.~\ref{sec:motivation}.
The influence of a short range spin-dependent interaction on the observable $R$ is explained and derived in Sec.~\ref{sec:ExoticForceOnR}.
Our result is compared to other current limits on the product $g_Sg_P$ in Sec.~\ref{sec:Comp}.

\section{Motivation}
\label{sec:motivation}

The investigation of CP violating processes is 
a major line of research in particle physics.
In contrast to the weak interaction, there is so far no evidence that the strong interaction violates CP symmetry.
The non-observation of an nEDM at current sensitivity levels constrains the CP violating term ({\it $\theta$-term}) in the Lagrangian of the strong interaction to be nine orders of magnitude smaller than naturally expected \cite{Pospelov2005}.
This fact is known as \textit{the strong CP problem} and a solution to it was proposed in \cite{peccei1977cp}, 
where the spontaneously broken Peccei-Quinn symmetry was introduced.

A new pseudoscalar boson emerges from this symmetry, the axion \cite{Wilzek1978,Weinberg1978}.
An intrinsic feature of the Peccei-Quinn model is a fixed relation between mass and interaction strength of the axion.
The originally assumed symmetry breaking scale (corresponding to the electroweak scale) was ruled out, leaving only higher energy scales possible.
For the axion one thus expects a small mass and a feeble interaction with other particles.
The possible mass of the axion is constrained by cosmology and astro-particle physics measurements to the so-called \textit{axion window} \cite{PDG14}.

\begin{figure}[h]
\centering
\includegraphics[width=\columnwidth]{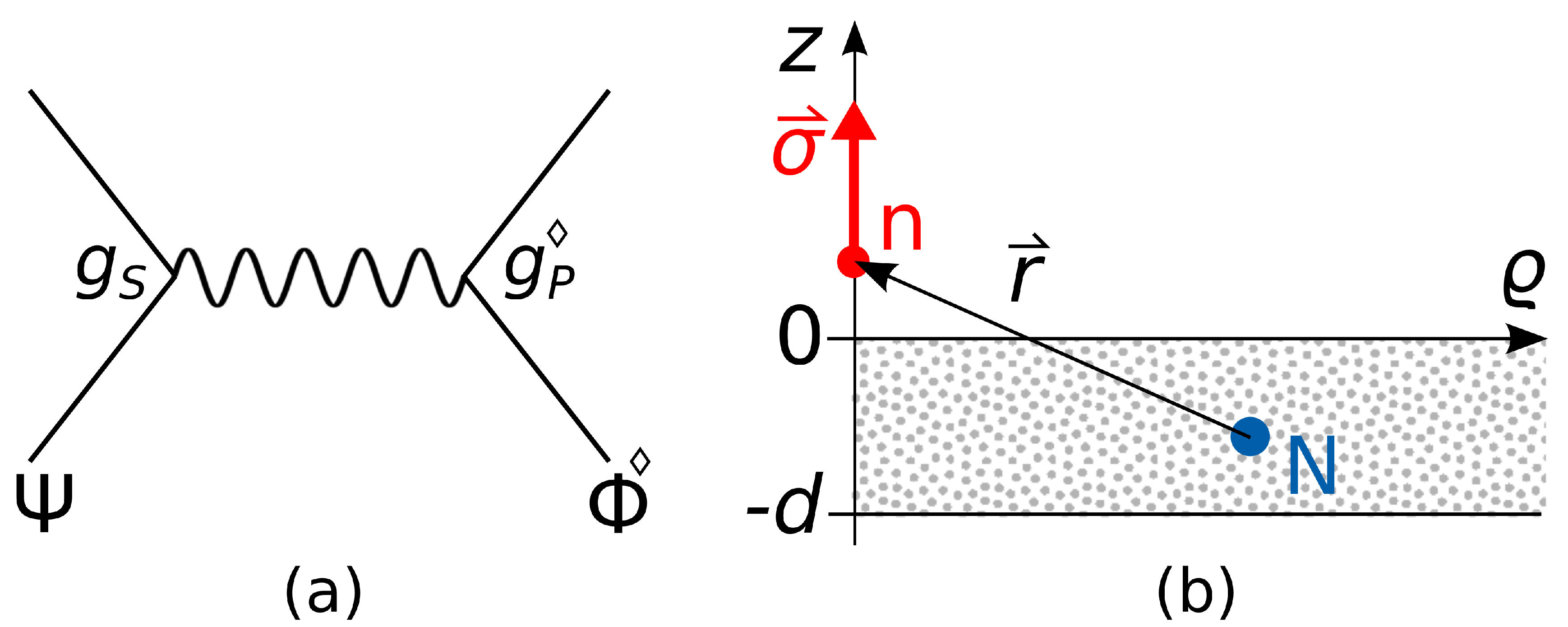}%
\caption{(a) Interaction diagram of a scalar-pseudoscalar coupling between particles $\Psi$ and $\Phi^{\diamond}$. 
$\Psi$ is unpolarized and interacts at the scalar vertex with the coupling constant $g_S$, whereas $\Phi^{\diamond}$ is polarized and interacts at the pseudoscalar vertex with the coupling constant $g_P^{\diamond}$.
The total interaction strength is proportional to the product $g_Sg_P^{\diamond}$.
(b) A polarized neutron n with spin $\boldsymbol{\sigma}$ interacts with an unpolarized nucleon N at distance $\boldsymbol{r}$ within bulk matter shaped as a plate of thickness $d$.
A view of the $\varrho$-$z$ plane in a cylindrical coordinate system $(\varrho,\phi,z)$ is shown.
}%
\label{fig:gSgP_bulkmatter}%
\end{figure}

A short range spin-dependent interaction which could be mediated by an axion
was proposed in \cite{MoodyWilczek84}.
There, three classes of interactions were presented, involving either $g_S^2$-, $g_Sg_P$-, or $g_P^2$-couplings, whereas $g_Sg_P$-couplings are considered of particular interest, since they violate CP symmetry.
A $g_Sg_P$-coupling diagram is shown in Fig.~\ref{fig:gSgP_bulkmatter}\,(a) and takes place between an unpolarized particle $\Psi$ (where \textit{unpolarized} means randomly polarized with respect to any quantization axis) and a polarized particle $\Phi^{\diamond}$.
The symbol $^{\diamond}$ is used to denote properties 
of the particle interacting 
at the pseudoscalar vertex with a strength proportional to the coupling constant $g_P^{\diamond}$ of the particle $\Phi^{\diamond}$.
The potential caused by such a $g_Sg_P^{\diamond}$-coupling between an unpolarized particle and a polarized particle with mass $m^{\diamond}$ and spin $\boldsymbol{\sigma}^{\diamond}$ 
is derived as \cite{MoodyWilczek84}:
\begin{equation} 
V(\boldsymbol{r})=g_S g_P^{\diamond}\frac{(\hbar c)^2}{8\pi m^{\diamond} c^2}
(\boldsymbol{\hat{\sigma}}^{\diamond}\cdot\boldsymbol{\hat{r}})
 \left(\frac{1}{r\lambda}+\frac{1}{r^2} \right)e^{\nicefrac{-r}{\lambda}},
\label{eq:NFpotential}
\end{equation} 
where $\boldsymbol{\hat{\sigma}}^{\diamond}$ is the unit vector of the spin, $\boldsymbol{\hat{r}}$ is the unit vector along the distance $\boldsymbol{r}$ between the particles, and $\lambda$ the interaction range.
The product $(\boldsymbol{\hat{\sigma}}^{\diamond}\cdot\boldsymbol{\hat{r}})$ also violates parity P and time reversal symmetry T. 

$g_Sg_P$-couplings can also be mediated by other hypothetical spin-zero particles which are generic to the axion and usually referred to as axion-like particles.
However, for these generic bosons no relation between mass and interaction strength is given, as compared to the genuine axion.
The origin of such particles can be symmetries other than Peccei-Quinn symmetry, which are broken at very high energies and often postulated in theories beyond the Standard Model of particle physics, such as e.g.~String Theory. 
Thus, both axions and axion-like particles, are intriguing dark matter candidates and beyond Standard Model physics probes \cite{antoniadis2011,RingwaldJaeckel2012,BakerAxionQuest2013}.

However, due to the non-observation of the nEDM a short range spin-dependent interaction mediated by an axion is constrained to $g_Sg_P<10^{-40}...10^{-34}$ \cite{RamseyMusolf2014}. 
On the other hand, if the force is mediated by an axion-like particle, $g_S$ and $g_P$ are not related to a specific symmetry breaking scale.
Thus, no significant constraint (i.e.~comparable to experimental sensitivity ranges) on $g_Sg_P$ can be deduced from current EDM limits \cite{RamseyMusolf2014} for the case of a generic boson being the interaction mediator.

Our measurement with ultracold neutrons is particularly sensitive to axion-like particles with a mass in the range of roughly 10\,meV to 100\,meV coupling to fermions. 
It also matches the mass range targeted by helioscopes such as CAST \cite{CAST2011} which would be sensitive to axion-like particles coupling to photons.

\section{The measurement with the nEDM apparatus}
\label{sec:ExoticForceOnR}

Spin-polarized ultracold neutrons of energies below 150\,neV are confined in a cylindrical storage chamber with vertical axis, height $H$=12\,cm, and inner diameter \o=47\,cm at the center of the nEDM apparatus.
A vertical magnetic holding field of $\sim$1\,\textmu T is applied.
The UCN spins precess for 180\,s and the precession frequency is inferred using Ramsey's method \cite{nEDMatPSI2011}.
The spins of polarized $^{199}$Hg atoms precess simultaneously in the same volume allowing to correct the Larmor precession frequency of the neutrons for magnetic field fluctuations.

We search for a signature of a spin-dependent interaction between polarized particles inside the storage chamber and the unpolarized wall of this chamber. 
This interaction can be described by the potential of Eq.~\ref{eq:NFpotential} and an example is shown in Fig.~\ref{fig:gSgP_bulkmatter}\,(b).
Integrating the interaction over all the nucleons present in uniform bulk matter results in an effective field normal to the surface. 

Since the potential is spin-dependent, it can also be regarded as a pseudomagnetic field $b$ which can affect the Larmor frequency of precessing spins.
For a symmetric setup with identical material for the bottom and top of the storage vessel, the field at the vessel surfaces points in opposite directions.
Therefore, we expect no shift in the Larmor frequency of the Hg atoms which sample the volume homogeneously.
However, UCNs have such low energies that they are significantly affected by gravity and their density increases towards the bottom of the storage vessel.
Thus, the vessel is inhomogeneously sampled 
and the effect of a pseudomagnetic field at the bottom of the chamber will not cancel out completely.
Depending on the sign of the vertical magnetic field, the precession 
frequency of the UCN spins will be increased or decreased. 
Consequently, $R$ will be shifted by a constant with opposite sign for the upward or downward oriented magnetic holding field $B_0$:
\begin{equation} 
R^{\uparrow\downarrow}=
\frac{\gamma_\mathrm{n}}{\gamma_\mathrm{Hg}}\left( 1\pm\frac{b}{B_0}\right),
\label{eq:b_on_R}
\end{equation} 
where the $\pm$ sign applies to the upward/downward oriented magnetic holding field, respectively.
The derivation of $R$ together with relevant systematic effects is detailed in \cite{pignol2014}.

\subsection{Derivation of the pseudomagnetic field}

Integrating  the potential given in Eq.~\ref{eq:NFpotential} over bulk matter, e.g.~a plate of thickness $d$ and nucleon number density $N$, results in a total potential $V_\mathrm{tot}$ at height $z$ above the surface of the plate
(cylindrical coordinates $\boldsymbol{r}=(\varrho,\phi, z)$ are used) \cite{Franke2013}:
\begin{align} 
V_\mathrm{tot}(z)&=
\int\limits_0^{-d} 
\int\limits_0^{2\pi} 
\int\limits_0^\infty 
NV(\boldsymbol{r})\,\ 
\varrho\,\mathrm{d}\varrho\, 
\mathrm{d}\phi\,
\mathrm{d}z=\\
&=g_Sg_P^{\diamond}\frac{\hbar^2N\lambda}{4m^{\diamond}}\left(1-e^{-\nicefrac{d}{\lambda}} \right) e^{-\nicefrac{z}{\lambda}}.
\label{eq:pseudoV}
\end{align} 
Since the interaction range $\lambda$ is much smaller than the chamber diameter, the integration over the radial component $\varrho$ is simplified to an infinite plane.
The pseudomagnetic field $b$ normal to the surface can be written as a function of \mbox{height $z$}:
\begin{equation} 
b(z)=\frac{2V_\mathrm{tot}(z)}{\gamma^{\diamond}\hbar}=
 g_S g_P^{\diamond}\frac{\hbar \lambda N}{2\gamma^{\diamond}m^{\diamond}} \left( 1-e^{-\nicefrac{d}{\lambda}}\right)
\left( e^{-\nicefrac{z}{\lambda}}\right),
\label{eq:pseudoB}
\end{equation}
where $\gamma^{\diamond}$ is the gyromagnetic ratio of the polarized particle.
This result agrees with the one derived in \cite{Zimmer2010physB}.

In order to calculate the field present in the nEDM setup $b_\mathrm{nedm}$, 
both the bottom and the top of the storage vessel have to be taken into account.
The vessel's inner horizontal surfaces are perpendicular to $\boldsymbol{z}$, with the bottom surface 
at $z=\nicefrac{-H}{2}$ and the top surface at $z=\nicefrac{+H}{2}$.
Using Eq.~\ref{eq:pseudoB} 
one finds
\begin{equation} 
b_\mathrm{nedm}(z)=b_\mathrm{bottom}~e^{-\frac{z+H/2}{\lambda}}-b_\mathrm{top}~e^{-\frac{-z+H/2}{\lambda}}.
\label{eq:pseudoBnedm}
\end{equation} 
Because of their inhomogeneous density distribution $\rho(z)$, the UCN experience an effective field given by
\begin{equation} 
b_\mathrm{UCN}=\int\limits_{-\frac{H}{2}}^{\frac{H}{2}} b_\mathrm{nedm}(z) \rho(z)\,\ \mathrm{d}z.
\label{eq:NFmagUCN}
\end{equation} 
The first order estimate for the UCN density distribution is 
\begin{equation} 
\rho(z)=\frac{1}{H}\left(1-\frac{12h}{H^2} z\right),
\label{eq:UCNdensity}
\end{equation} 
where $h$=2.35(5)\,mm is the measured  center-of-mass offset between UCN and Hg atom distributions \cite{pignol2014}.

Since the pseudomagnetic field is expected to be of short interaction range $\lambda$ given by the limits 
which have already been imposed on $g_Sg_P$, the UCN density distribution can be approximated 
with a constant value within a distance $\sim$$\lambda$ to the bottom and top surfaces.
As a consequence, Eq.~\ref{eq:NFmagUCN} can be simplified in the following way:
\begin{equation} 
b_\mathrm{UCN} \approx \int\limits_{-\frac{H}{2}}^{\frac{H}{2}}
\left(
\rho_\mathrm{bottom}\,b_\mathrm{bottom}\,e^{-\frac{z+H/2}{\lambda}} - 
\rho_\mathrm{top}\,b_\mathrm{top}\,e^{-\frac{-z+H/2}{\lambda}}
\right)
\, \mathrm{d}z,
\label{eq:NFmagUCNsimple}
\end{equation} 
where $\rho_\mathrm{bottom}=\rho(\nicefrac{-H}{2})$ and $\rho_\mathrm{top}=\rho(\nicefrac{+H}{2})$. 
Thus the integral can be solved analytically and 
the scalar and pseudoscalar coupling constants can be isolated as follows:
\begin{equation} 
g_Sg_P^{\diamond}=b_\mathrm{UCN}\frac{H^2\gamma^{\diamond}m^{\diamond}}{6\hbar h N \lambda^2}
 \left( 1-e^{-\nicefrac{H}{\lambda}} \right)^{-1}
\left( 1-e^{-\nicefrac{d}{\lambda}} \right)^{-1}.
\label{eq:gSgPisolated}
\end{equation} 
The bottom and top of the UCN storage vessel are made of aluminum
and have a thickness of $d=2.5$\,cm each.
We use $N\equiv N_\mathrm{Al}=1.62\cdot10^{24}\,\mathrm{cm}^{-3}$ and $\gamma^{\diamond}\equiv \gamma_\mathrm{n}=2\pi\cdot29.1646943$\,MHz/T \cite{Mohr2012}.
The surface of the aluminum plates is coated with diamond-like carbon in order to improve neutron storage properties.
The coating thickness is below 3\,\textmu m and the density of diamond-like carbon is similar to that of aluminum.
Thus the coating is disregarded in the calculation but for the fact that we restrict the validity of our derived limit to 
$\lambda>1$\,\textmu m.

The center-of-mass offset $h$ contributes to the denominator of Eq.~\ref{eq:gSgPisolated} and depends on the energy spectrum of the UCN. Hence, a change in the energy spectrum, or also a different vessel height $H$, will influence the sensitivity of our apparatus to a pseudomagnetic field.

In \cite{pignol2014}, two independent measurements for $R$ are presented, one for each direction of the magnetic holding field (upwards and downwards, respectively):
\begin{eqnarray}
R^{\uparrow}&=&3.8424583(26)\label{eq:R1} \\
R^{\downarrow}&=&3.8424562(30).
\label{eq:R2}
\end{eqnarray} 
The sensitivity of the interpretation presented here depends explicitly on the
uncertainties given in Eqs.~\ref{eq:R1} and \ref{eq:R2}. 
Statistical and systematical contributions to these uncertainties are 
discussed in \cite{pignol2014}.

From the difference of $R^{\uparrow}$ and $R^{\downarrow}$ (see Eq.~\ref{eq:b_on_R}) we can extract 
\begin{equation}
b_\mathrm{UCN}=\left( R^{\uparrow}-R^{\downarrow}\right)\frac{\gamma_\mathrm{Hg}}{\gamma_\mathrm{n}}
\frac{B_0}{2}= (0.28\pm 0.53)\,\mathrm{pT}.
\label{eq:pseudoBmeasured}
\end{equation}
%

\section{Result and comparison to other experiments}
\label{sec:Comp}

\begin{figure}[h] 
\centering
   \includegraphics[width=\columnwidth] {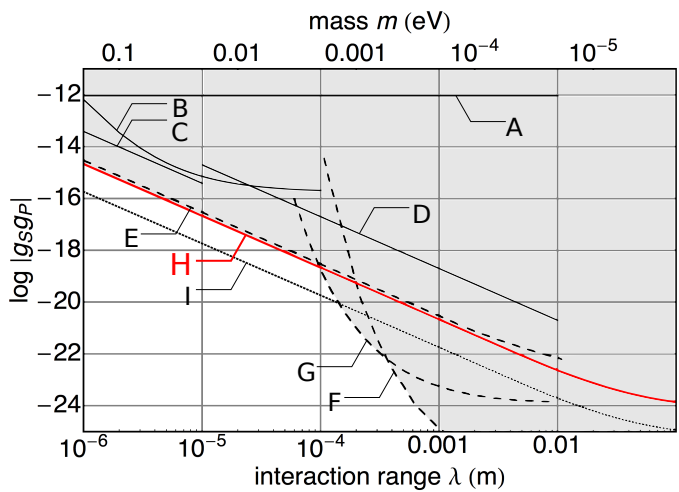}
  \caption[Overview of current limits on the product of scalar and pseudoscalar coupling constants $g_Sg_P$.]
  {Overview of current limits on the product of scalar and pseudoscalar coupling constants $g_Sg_P$ as function of the interaction range $\lambda$ of a short range spin-dependent force at 95\,\% confidence level.
  On the top, the corresponding mass range of the mediating particle, i.e.~axion or axion-like particle, is shown.
  The shaded region is excluded by different experiments.
  Solid line limits were obtained using cold or ultracold neutrons.
  Dashed line limits were obtained using $^3$He, $^{129}$Xe, or $^{131}$Xe precession experiments.
  A \cite{Voronin2009}; B \cite{JenkePRL2014}, assuming an attractive interaction; C \cite{Serebrov2009Axion};
  D \cite{Serebrov2010Axion}; E \cite{Petukhov2010}; 
  F \cite{tullney2013}; G \cite{Bulatowicz2013}; 
  and H (red) this work. The line I (dotted) depicts the achievable limit by a simple modification of our apparatus (see text). 
  }
  \label{fig:gSgPErrorBudget}
\end{figure}

Using Eq.~\ref{eq:gSgPisolated}, the measured pseudomagnetic field $b_\mathrm{UCN}$ can be converted to a 95\,\% confidence level limit on $g_Sg_P$
\begin{equation}
g_Sg_P \lambda^2<2.2\cdot10^{-27}\,\mathrm{m}^2
\label{eq:gSgP95CL}
\end{equation}
for 1\,\textmu$\mathrm{m}<\lambda<5$\,mm.
At the upper end of this range, 
the last factor in Eq.~\ref{eq:gSgPisolated} departs from $\sim$1 and the relation $g_Sg_P\propto 1/\lambda^2$ is not fulfilled anymore.
As a consequence the sensitivity of our experiment to $g_Sg_P$ decreases which results in a flattening of the limit in the $g_Sg_P$ vs.~$\lambda$-plane (see Fig.~\ref{fig:gSgPErrorBudget}).
The lower end of this range is constrained by the wavelength of ultracold neutrons and affected by surface properties such as coating, roughness, etc.

Since we investigated an interaction between unpolarized nucleons and polarized neutrons, we can state that we 
probed the scalar coupling constant generally valid for nucleons $g_S\equiv g_S^{\mathrm{N}}$ and the pseudoscalar coupling constant specific to the neutron $g_P^{\diamond}\equiv g_P^{\mathrm{n}}$.

Figure \ref{fig:gSgPErrorBudget} compares our limit on $g_Sg_P$ to results from other experiments.
It covers the interaction range of \mbox{1\,\textmu$\mathrm{m}<\lambda<0.1\,\mathrm{m}$}, which is not yet strongly excluded by astrophysical or cosmological constraints.

Experiments using free neutrons are depicted by solid lines.
Experiments with precessing atoms, 
such as e.g.~$^3$He, $^{129}$Xe, or $^{131}$Xe, are depicted by dashed lines.
According to the Schmidt Model \cite{schmidt1937}, polarized atoms of odd isotopes (with one unpaired nucleon) can roughly be considered as a probe for the magnetic properties of this unpaired nucleon, regardless of the other constituents of the nucleus.
Under this assumption,
both types of experiments probe $g_S^{\mathrm{N}}g_P^{\mathrm{n}}$. 
While these approaches are complementary, the direct neutron measurements are model independent.
The most stringent limits for $\lambda>10^{-4}$\,m have been 
imposed recently in \cite{tullney2013} and \cite{Bulatowicz2013} 
(curves labelled F and G in Fig.~\ref{fig:gSgPErrorBudget}, respectively)
which improved the recent limits from \cite{SnowChu2013}.
For shorter interaction ranges, the most stringent limit was given in \cite{Petukhov2010} (E), where relaxation of spin polarized $^3$He gas was investigated.
This limit has been model independently confirmed and slightly improved by the measurements presented in this work (H).
Limits derived from experiments with 
neutrons (A,B,C,D) \cite{Voronin2009,JenkePRL2014,Serebrov2009Axion,Serebrov2010Axion}
were improved by one order of magnitude for $\lambda<10^{-5}$ and by two orders of magnitude for $\lambda>10^{-5}$.
In \cite{Raffelt2012} a stronger but indirect limit on $g_Sg_P$ was imposed
by combining laboratory results with stellar energy loss arguments.
Such limits might be reached with dedicated future laboratory searches e.g.~proposed in \cite{Arvanitaki2014}. 

Already our present result constitutes a new direct limit on $g_Sg_P$.
Replacing in our experiment the central vessel bottom and top with copper,
a material with higher density and good UCN reflecting surface properties,
would result in a sensitivity gain of $\sim$3, corresponding to the density ratio
between copper and aluminum.
Replacing either only the bottom or top 
would create a true asymmetric potential and
increase the sensitivity by one order of magnitude \cite{Franke2013}.
The consequently achievable limit depicted by the dotted curve (I) in Fig.~\ref{fig:gSgPErrorBudget}
would be an important 
contribution to 
reduce the allowed parameter spaces of beyond Standard Model theories.

\section{Acknowledgements}
We are grateful to the PSI staff (the accelerator operating
team and the BSQ group) for providing excellent running conditions
and acknowledge the outstanding support of M.~Meier and F.~Burri. 
Support by the Swiss National Science Foundation
Projects 200020-144473 (PSI), 200021-126562 (PSI), 200020-149211
(ETH) and 200020-140421 (Fribourg) is gratefully acknowledged.
The LPC Caen and the LPSC Grenoble acknowledge the support
of the French Agence Nationale de la Recherche (ANR) under
reference ANR-09–BLAN-0046. 
The Polish collaborators acknowledge
the National Science Centre, Poland, for the grant
No.~UMO-2012/04/M/ST2/00556
and the support by the Foundation for Polish Science - MPD program, co-financed by the European Union within the European Regional Development Fund.
This work was partly supported
by the Fund for Scientific Research Flanders (FWO), and
Project GOA/2010/10 of the KU Leuven. The original apparatus
was funded by grants from the UK's PPARC.

\end{document}